\documentclass{iau}

\usepackage{amsmath}
\usepackage{graphicx}
\usepackage{multirow}

\begin{document}

\lefttitle{G. C. Van de Steene et al. }
\righttitle{VLTP Sakurai's object: [WC] star in a new bipolar nebula}

\journaltitle{Planetary Nebulae: a Universal Toolbox in the Era of Precision Astrophysics}
%\jnlPage{1}{7}
\jnlDoiYr{2023}
\doival{10.1017/xxxxx}
\volno{384}

\aopheadtitle{Proceedings IAU Symposium}
\editors{O. De Marco, A. Zijlstra, R. Szczerba, eds.}
 
\title{Sakurai's object: a [WC] star in a new bipolar nebula after a VLTP}

\author{G. C. Van de Steene$^{1}$ , P. A. M. van Hoof$^{1}$, S.
  Kimeswenger$^{2,3}$, M. Hajduk$^{4}$ ,D. Tafoya $^{5}$,
  J. A. Toalá$^{6}$, A. A.  Zijlstra$^{7}$ , D. Barría$^{8}$
}

\affiliation{
 $^{1}$Department of Astronomy \& Astrophysics, Royal Observatory of
   Belgium, Ringlaan 3, 1180 Brussels, Belgium; 
$^{2}$Universit{\"a}t Innsbruck, Institut f{\"u}r Astro- und
   Teilchenphysik, Technikerstr. 25, 6020 Innsbruck, Austria; 
$^{3}$Instituto de Astronom{\'i}a, Universidad Cat\'olica del Norte, Av. Angamos 0610, Antofagasta, Chile; 
$^{4}$Department of Geodesy, Faculty of Geoengineering, University of Warmia and Mazury, ul. Oczapowskiego 2, 10-719 Olsztyn, Poland; 
 $^{5}$Department of Space, Earth and Environment, Chalmers University of Technology, Onsala Space Observatory, 439~92 Onsala, Sweden; 
$^{6}$Instituto de Radioastronom\'{i}a y Astrof\'{i}sica, UNAM, Ant. Carretera a P\'{a}tzcuaro 8701, Ex-Hda. San Jos\'{e} de la Huerta, Morelia 58089, Mich., Mexico; 
$^{7}$Jodrell Bank Centre for Astrophysics, Alan Turing Building,
  University of Manchester, Manchester, M13 9PL, UK; 
$^{8}$Facultad de Ingeniería y Arquitectura, Universidad Central de Chile, Av. Francisco de Aguirre 0405, La Serena, Coquimbo, Chile }

\begin{abstract}
  
  Optical spectra of the Very Late Thermal Pulse (VLTP) object V4334\,Sgr have shown a rapidly changing spectrum resulting from shocks in the outflow, which 
  created a new bipolar nebula inside the old nebula. We see C\,II and C\,III emission lines
  emerging typical of a [WC\,11-10]-type star.  The strong increase of  [O\,III] and [S\,III] 
  emission lines indicate the possible onset of photoionisation in the new ejecta.

\end{abstract}


\maketitle

\section{Introduction}

V4334 Sgr (Sakurai's object) is the central star (CS) of an old
Planetary Nebula (PN) that underwent a Very Late Thermal Pulse (VLTP)
around 1996.  The VLTP forced the star to briefly
track back to the AGB and experience a second PN formation stage.
High resolution ALMA continuum observations have shown the
presence of a new bipolar PN inside the old one \citep{Tafoya23}. We
have been observing this object since 2006 to improve our understanding 
of PN formation and evolution.

\section{Optical Monitoring}

The first emission lines from the newly ejected gas were discovered in
1998 \citep{Eyres99} and 2001 \citep{Kerber02}.  We have been monitoring the changes in the
optical emission line spectrum  since 2006 using spectra obtained with
FORS1\&2 at ESO-VLT \citep{Reichel22,vanHoof07,VandeSteene17}. The goal of this monitoring program has been to
derive the stellar temperature as a function of time in order to
compare with stellar evolutionary models \citep{Bertolami06,Herwig14}, which predict a fast reheating to 80\,kK on a
timescale of decades \citep{Hadjuk05}. Such fast evolution would
cause the ejecta to be photoionised rapidly, resulting in observable,
dramatic changes in the emission line spectrum.

\section{Evolution of line fluxes}

\begin{figure}
\begin{center}
\includegraphics[width=8.4cm]{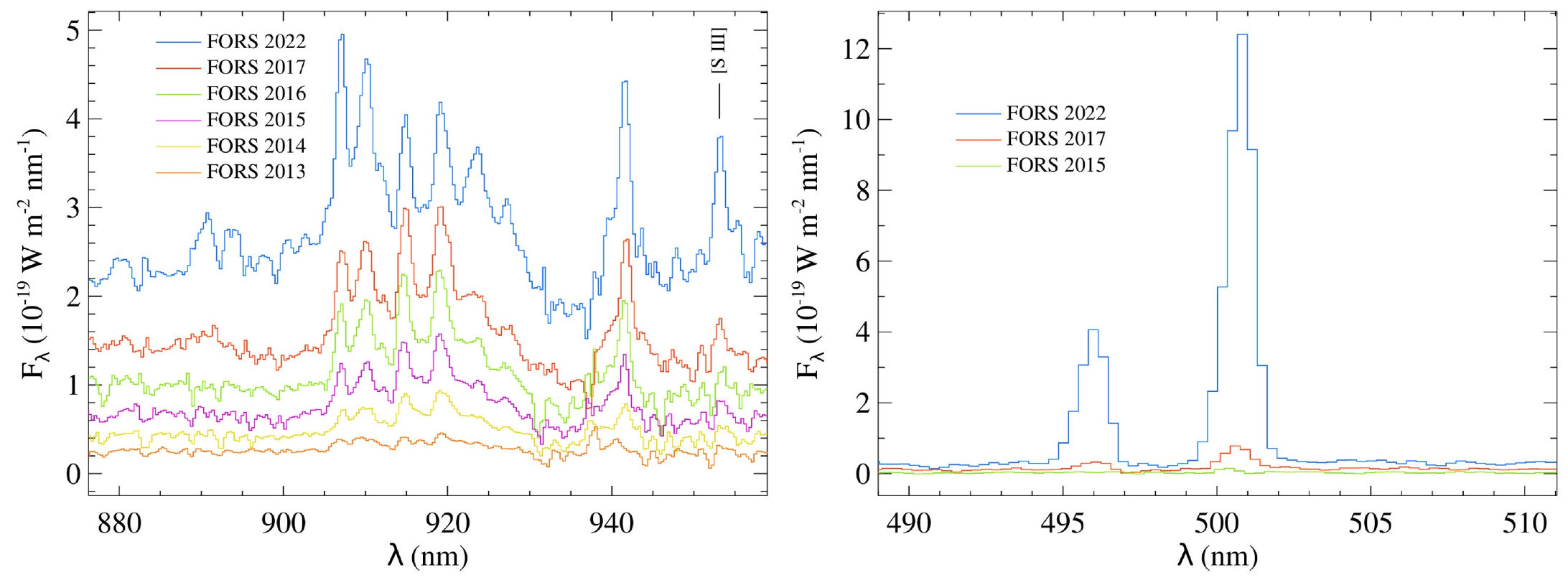}
\includegraphics[width=4.2cm]{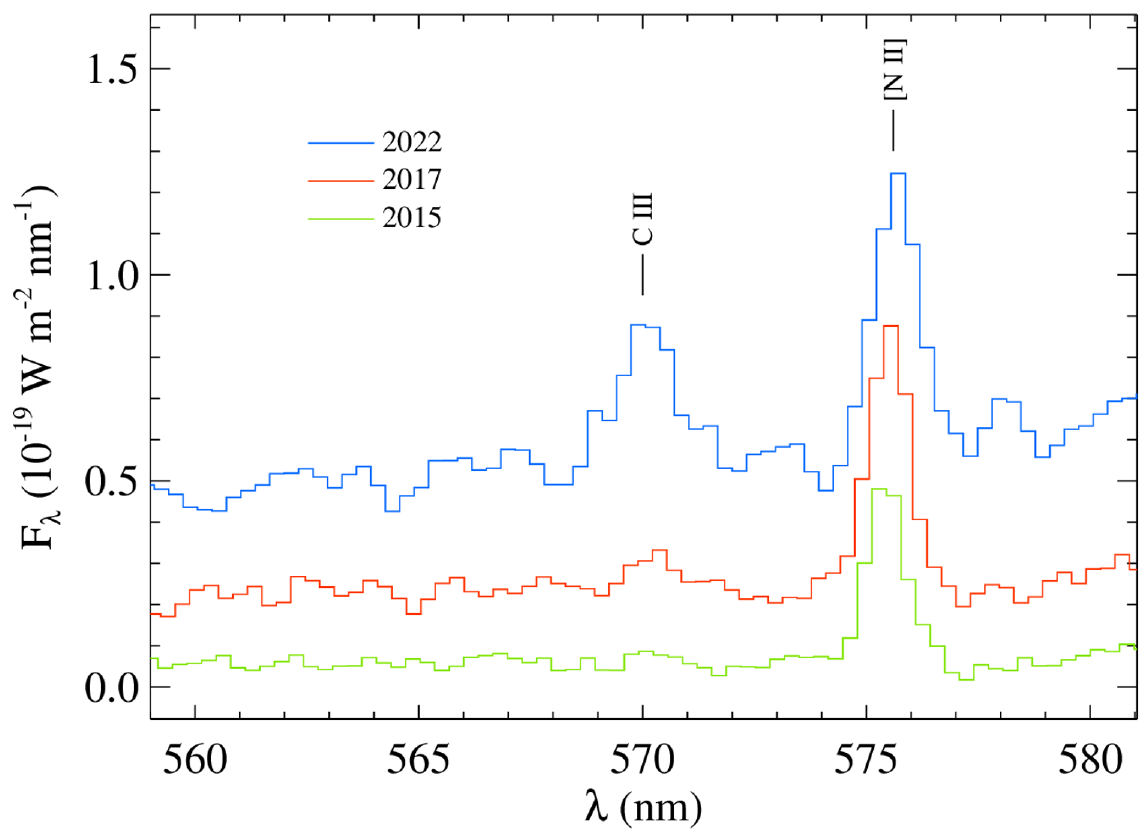}
\end{center}
\caption{The emergence of WR lines since 2013 (left), the increase in [O\,III] 
            emission lines (middle), and the emergence of the C\,III$\lambda$569.6\,nm line (right).}
\label{lines} 
\end{figure}

From 2001 to 2017 the data show a rapidly evolving spectrum that is
completely produced by shocks. These shocks have been the result
of the hydrodynamic shaping of the new bipolar nebula forming inside the old nebula.
In \cite{vanHoof07} we argued for the existence of
a shock that occurred around 1998 and dissipated shortly afterwards
based on the optical decline of the line flux values.
In 2008 the spectrum changed dramatically. The [OII]\,$\lambda$732.5\,nm
lines increased almost tenfold in strength, and weak He\,I emission
appeared. This sudden increase in excitation would have been caused by
a second shock. However, unlike the first shock, there has been a
gradual increase in line strengths since then \citep{vanHoof18}. The
evolution of the low-ionization forbidden lines is still dominated by
these shocks in the most recent spectra. The emergence of
a substantial stellar wind may have been the cause of the second
shock, which is associated with the bipolar bow-shock features
discovered in \cite{Hinkle14}. In 2014 there was another dramatic change in the spectrum with the
emergence of a complex of emission lines mostly between $\lambda$905.0
and $\lambda$945\,nm (Fig.\ref{lines},left). Many are identified as
carbon lines \citep{Williams21}. The flux evolution of these lines
differs markedly from the shock excited helium and forbidden (nebular)
lines. This points to a different origin and we assume that these
lines are formed in the wind of the central star and are in fact WR
features. The emergence of C\,III lines in the most recent spectra
would confirm this assumption (Fig.\ref{lines}, right). The presence of C\,II and
C\,III lines, but the absence of C\,IV lines indicate a spectral type
[WC\,11-10] \citep{Acker03}. This is the first sign of reappearance of
the central star after the VLTP.  In 2017 the spectra showed the first
tentative signs of photoionisation (Fig.\ref{lines}, middle) with a sudden
increase in the flux of the [O\,III]\,$\lambda$495.9 and
$\lambda$500.7\,nm lines compared to 2015. This trend has continued in
the spectra of 2022 where these lines showed another dramatic increase
in strength. For the first time also the [S\,III]$\lambda$953.1\,nm
line appeared (Fig.\ref{lines}, left). Its $\lambda$906.9\,nm counterpart
is blended. However, radio continuum observations have shown only a
slight increase of flux density (Hadjuk et al. this volume). It needs to be confirmed
whether photoionisation of the new ejecta by the central star has
started or this increase is due to strenghtening of the shock.

\newcommand*\aap{A\&A}
\newcommand*\apj{ApJ}
\newcommand*\apjl{ApJL}
\newcommand*\aj{AJ}
\newcommand*\mnras{MNRAS}
\newcommand*\iaucirc{IAU~Circ.}
\newcommand*\iaus{IAUS}

\end{document}